\documentclass[11pt]{article}
\usepackage[normalem]{ulem}
\pdfoutput=1 

\usepackage{jheppub} 
\usepackage{url}
\usepackage{caption}
\usepackage{subcaption}
\usepackage{extarrows}

\usepackage[latin9]{inputenc}
\usepackage{float}
\usepackage{amsmath}
\usepackage{amssymb}
\usepackage{graphicx}
\usepackage{esint}
\usepackage{hyperref}
\usepackage{comment}
\usepackage{color}
\usepackage{microtype}
\usepackage{cleveref}
\usepackage{breakurl}
\usepackage{bbm}
\newcommand{\be}{\begin{equation}}
\newcommand{\ee}{\end{equation}}
\newcommand{\ben}{\begin{displaymath}}
\newcommand{\een}{\end{displaymath}}
\newcommand{\bea}{\begin{eqnarray}}
\newcommand{\eea}{\end{eqnarray}}
\def\K{K{\"a}hler }
   \newcommand{\rf}[1]{(\ref{#1})}
\newcommand{\vp}{\varphi}

\def\be{\begin{equation}}
\def\ee{\end{equation}}
\def\bea{\begin{eqnarray}}
\def\eea{\end{eqnarray}}
\def\ba{\begin{array}}
\def\ea{\end{array}}
\def\bit{\begin{itemize}}
\def\eit{\end{itemize}}

\def\a{\alpha}

\def\vp{\varphi}

 \makeatletter

\allowdisplaybreaks

\makeatother

\makeatletter
\DeclareRobustCommand{\rcite}[1]{%
  \rcite@aux#1,\@nil{#1}%
}
\def\rcite@aux#1,#2\@nil#3{%
  \if\relax#2\relax
    Ref.~\cite{#3}%
  \else
    Refs.~\cite{#3}%
  \fi
}
\makeatother

\hypersetup{
    colorlinks = true,
    citecolor = {blue},
    linkcolor = {blue},
    urlcolor = {blue},
}

\def\rme{{\rm e}}

\newcommand{\lp}{\left(}

\newcommand{\rp}{\right)}

\title{\rm { \LARGE \bf    Streamlined Supergravity  }    }

\author{Renata Kallosh and Andrei Linde}

\affiliation{Leinweber Institute for Theoretical Physics at Stanford, 382 Via Pueblo, Stanford, CA 94305, USA}
\emailAdd{kallosh@stanford.edu}
\emailAdd{alinde@stanford.edu}

\abstract{ The textbook N=1 supergravity has an F-term  potential depending on a superpotential $W(z_i)$ and a \K potential $K(z^i, \bar z^{\bar i})$, with the scalar potential  $V(z^i, \bar z^{\bar i})=e^K (|DW|^2 - 3 |W|^2)$. In this approach, it is not always easy to find the potential $V(z^i, \bar z^{\bar i})$ with the required properties. We show that in supergravity with a nilpotent superfield and {\it with any  \K potential}  $K(z^i, \bar z^{\bar i} )$ one can obtain {\it any desired potential}  $V(z^i, \bar z^{\bar i})$ by a proper choice of the \K metric of the nilpotent superfield.  This construction is particularly suitable for cosmological and particle physics applications, which may require maximal freedom in the choice of kinetic terms and scalar potentials. 
}

\begin{document}

\maketitle

\parskip 7 pt

\section{Introduction} 
Supergravity textbooks \cite{Wess:1992cp, Freedman:2012zz} and a more recent ``Simple Supergravity'' review \cite{DallAgata:2022nfr} describe supergravities with linearly realized supersymmetry. 
The models with non-linearly realized supersymmetry have been actively studied over the last couple of decades, particularly in the context of cosmological applications of supergravity. These are best described as supergravity models that, in addition to unconstrained physical chiral multiplets $(Z^{i}, \bar Z^{\bar i})$, contain one nilpotent chiral multiplet $(S, \bar S)$ with $S^2=0$. This is a supergravity version of the anti-D3 brane uplifting of the potential in the KKLT model \cite{Kachru:2003aw}, which is required to produce de Sitter vacua in string theory. Early descriptions of these models for inflation and dark energy can be found in \cite{Ferrara:2014kva} and in a more recent review \cite{Antoniadis:2024hvw}.

The presence of a nilpotent superfield made it possible to construct pure de Sitter supergravity \cite{Bergshoeff:2015tra,Hasegawa:2015bza}, to stabilize all scalars but one \cite{Ferrara:2014kva,Carrasco:2015rva}, and to develop a large class of inflationary models currently favored by the data. 

An important advantage of including the nilpotent multiplet, in addition to standard chiral multiplets, in the theory was emphasized in \cite{Kallosh:2015sea}. The potential of these models at the vanishing scalar $s$ of the nilpotent multiplet has the form
\be
V= e^K ( |F|^2 + |DW|^2 - 3 |W|^2) \ ,
\ee
where $|DW|^2 - 3 |W|^2$  is the standard chiral superfield contributions to the potential. Meanwhile, the term $ |F|^2$ is defined in eq. (2.6) in \cite{Kallosh:2015sea}, so that
\be
V|_{s=0}= e^K (F_s K^{s\bar s} \bar F_{\bar s} + D_i W K^{i\bar k} \bar D_{\bar k} \bar W - 3 W\bar W) \ .
\label{VTMF}\ee
Here $D_i W$ is a \K covariant derivative of the superpotential over the scalar $z^i$ of the physical chiral multiplet $Z^i$, 
$F_s= D_s W$ where $D_s W$ is a \K covariant derivative of the superpotential over the scalar $s$ of the nilpotent chiral superfield multiplet $S$.  Also,  $K^{s\bar s}$ is the inverse \K metric of the scalar from a nilpotent superfield, and  $K^{i\bar k}$ is the inverse \K metric of the physical scalars. Eq. \rf{VTMF} is valid under assumption made in \cite{Kallosh:2015sea} that 
$K_{s}|_{s=0}=0$, in particular $K_{s\bar k}=0$. 

Important steps in developing supergravity with one nilpotent multiplet in addition to standard chiral multiplets where made over the years, see for example  \cite{Antoniadis:2014oya,DallAgata:2014qsj,Dudas:2015eha,Kallosh:2015tea,Kallosh:2015pho,Ferrara:2016een,McDonough:2016der,Freedman:2017obq,Kallosh:2017wnt,Achucarro:2017ing,Yamada:2018nsk,Linde:2018hmx,Kallosh:2022vha,Kallosh:2025jsb}. A related new construction called ``Liberated Supergravity'' was developed in \cite{Farakos:2018sgq} based on linearly realized off-shell supersymmetry and  \K- Weyl invariance.

 \K manifolds with bisectional curvature  along the $S=0$ plane  in models with  heavy stabilizer field $S$ were studied in \cite{Kallosh:2010xz}, and with a  nilpotent superfield $S^2=0$ in \cite{Kallosh:2017wnt}: 
 \be
R^{\rm bisec}=-R_{ z\, \bar z\, s\, \bar s}K^{ z\, \bar z} K^{s\bar s}\, , \qquad R_{ z\, \bar z\, s\, \bar s}= \partial_z \partial _{\bar z} K_{\bar s s} + K^{s\bar s} \partial_z K_{s\bar s} \partial _{\bar z} K_{s\bar s}  \ .
\ee
 \K potentials of the form $K_{s\bar s}(z, \bar z) S\bar S$  have a bisectional curvature iff $K_{s\bar s}$ depends on physical moduli $(z, \bar z)$.   It was shown in \cite{Kallosh:2010xz} that terms involving the bisectional curvature play an important role in stabilizing moduli during inflation and in the properties of their inflationary fluctuations. Cosmological supergravity models with \K potentials mixing the stabilizer with a chiral sector $(z, \bar z)$ were proposed in \cite{Kallosh:2010xz}, and the ones 
 mixing a nilpotent superfield with a chiral sector were proposed in \cite{McDonough:2016der,Kallosh:2017wnt}.
 
  {\it Now we take the main step in constructing streamlined supergravity: we express the nilpotent superfield \K metric via an arbitrary potential $V$ depending on the physical scalars, and an arbitrary \K potential $K$ of the physical scalars, using eq. \rf{VTMF}. We   bring this equation from \cite{Kallosh:2015sea}  to the form}
\be\boxed{
K^{s\bar s} =  {e^{-K } V - D_i W K^{i\bar k} \bar D_{\bar k} \bar W - 3 W\bar W\over F_s \bar F_{\bar s}} \ .}
\label{solution1}\ee
We have thus multiplied eq. \rf{VTMF} by ${e^{-K}\over  F_s \bar F_{\bar s}}$ and moved $K^{s\bar s}$ to the left and ${e^{-K} V \over  F_s \bar F_{\bar s}}$ to the right. The resulting eq. \rf{solution1} is now valid for any $V(z^i, \bar z^{\bar i}) $ and $K(z^i, \bar z^{\bar i} )$ depending on physical multiplets, same as the original equation \rf{VTMF}.

Many examples of supergravities with nilpotent multiplets presented before in \cite{Kallosh:2015tea,Kallosh:2015pho,Ferrara:2016een,McDonough:2016der,Freedman:2017obq,Kallosh:2017wnt,Achucarro:2017ing,Yamada:2018nsk,Linde:2018hmx,Kallosh:2022vha,Kallosh:2025jsb},  see also \cite{Kallosh:2022ggf,Braglia:2022phb,Kallosh:2024ymt,Carrasco:2025rud}, were restricted either to potentials at the inflationary trajectory, such as $V|_{z=\bar z}$, or to specific choices of the \K potentials of physical scalars. Now we have a general case of $V(z^i, \bar z^{\bar i}) $ and $K(z^i, \bar z^{\bar i} )$, explaining {\it a posteriori} why things were working so well in various examples.

We will present below a simple version of supergravity defined in Eqs. \rf{VTMF}, \rf{solution1} which achieves the goal of developing consistent supergravity with a nilpotent superfield, such that {\it for any choice} of the \K potential of the chiral superfields $K(z^i, \bar z^{\bar i} )$ one can obtain {\it any desired potential}  $V(z^i, \bar z^{\bar i})$ by a proper choice of the \K metric $K_{s\bar s}(z^i, \bar z^{\bar i})$ of the nilpotent superfield.

This is very different from the textbook N=1 supergravity with a potential depending on a superpotential $W(z_i)$ and \K potential $K(z^i, \bar z^{\bar i})$ of the form  $V_{\rm standard}(z^i, \bar z^{\bar i})=e^K (|DW|^2 - 3 |W|^2)$, which often makes it  difficult to obtain the desired potential  $V(z^i, \bar z^{\bar i})$ by a proper choice of $W(z_i)$ and $K(z^i, \bar z^{\bar i})$.

\section {Construction of streamlined supergravity with  a nilpotent superfield}
Streamlining, by definition, means making something more efficient and simpler by removing unnecessary parts.  Note that the nilpotent sector acts behind the scenes; our only goal is to find a simple and efficient way to construct the most general scalar potentials and \K potentials in the physical sector. In our case, it
involves  {\it two simplifications in the nilpotent sector of the theory}.

\begin{enumerate}
  \item {\it Streamlining the dependence of the  \K potential  on the nilpotent superfield  }

This step  was already taken in \cite{Kallosh:2015sea,Kallosh:2015tea}, where the  \K potential was taken in the form
\be
K(z^i, \bar z^{\bar i},  s, \bar s)=K(z^i, \bar z^{\bar i} )+ K_{s\bar s} (z^i, \bar z^{\bar i} ) \, s \,\bar s \ .
\label{K}\ee
A more general case would be the one where we add the $z, \bar z$-dependent terms that are linear in $s$ and $\bar s$. The result is known \cite{Schillo:2015ssx},  supergravity is more complicated. However, for our purposes, such complications are not warranted: we can achieve our main goal without introducing these additional terms in the nilpotent sector of the theory.
 
  \item {\it  Streamlining the dependence of the superpotential $W$ on the nilpotent superfield, and making it $z$-independent}
  
We will use  a simple superpotential
  \be
W= W_0 +   s \, F_s \    .
\label{W}\ee
\end{enumerate}
The constant $W_0$ defines the gravitino mass, whereas the constant $F_s$ is a non-vanishing value of the $D_s W|_{s=0}= F_s$.  We keep here a simple \K potential  \rf{K} and a simple superpotential  \rf{W},  but our \K potential of physical scalars $K(z^i, \bar z^{\bar i} )$ is arbitrary. This choice of the simple $s$-sector is convenient and sufficient to construct supergravity with any desired potentials $V(z^i, \bar z^{\bar i})$.  Our choice of \rf{W} simplifies the general expression for 
$F_s= D_s W $ used in  \cite{Kallosh:2015sea} and makes it possible to use a constant $F_s$ instead of a function $F_s(z^i)$. 

 The original Volkov-Akulov theory of nonlinearly realized global supersymmetry \cite{Volkov:1972jx} did not rely on models with linearly realized supersymmetry with unconstrained chiral multiplets, since the Volkov-Akulov model was discovered earlier.  
In case of global supersymmetry we have  $n$ chiral superfields $Z^i= z^i + \sqrt 2\,  \theta \psi^i_z + \theta^2 f_z^i$ and one nilpotent superfield $S$, where  $S= s+ \sqrt 2\, \theta \psi^s + \theta^2 f_s$ and $S^2=0$. The nilpotency condition is resolved as follows, for $f_s\neq 0$
\cite{Casalbuoni:1988xh,Komargodski:2009rz},
\be
 s= {(\psi^s)^2\over 2 f_s} \ .
\label{casa}\ee
which brings the original theory with linear supersymmetry to a Volkov-Akulov type model with a non-linearly realized supersymmetry.
It was observed in \cite{Casalbuoni:1988xh} that the nilpotency condition $S^2=0$ is achieved in the model with linear supersymmetry in the limit of infinite scalar mass. This limit is performed by sending the vacuum curvature of the \K manifold to infinity. 
The nilpotency constraint \rf{casa} means that we have to impose a condition $s=0$ in the bosonic part of the action.  

In the case of local supersymmetry, the resolution of the nilpotency condition is a generalization of the one in global supersymmetry. 
Superconformal theory with a nilpotent multiplet was studied in \cite{Kallosh:2015sea,Kallosh:2015tea}. Transition to supergravity followed the steps described in \cite{Freedman:2012zz} for usual chiral and vector multiplets, and special attention was given to the presence of the nilpotent chiral multiplet. As a result, 
the resolution of the nilpotency condition was established in   \cite{Kallosh:2015sea,Kallosh:2015tea}. It takes the following form:
\be
 s= {(\psi^s)^2\over 2  F} \, , \qquad {\rm where} \quad  F  = e^{K/2} K^{s\bar s} \, \big(-\overline{D_s W}+ \dots\big) \ .
\label{F}\ee
 Here, $\dots$ denotes various terms containing fermions.
We will provide further details in the Appendix, describing the complete supergravity action, including scalars, fermions, and vectors. In supergravity  in \cite{Kallosh:2015sea} and here in eq. \rf{VTMF} in absence of fermions
\be
\bar F_{\bar s}=\overline{D_s W}= - e^{-K/2} K_{s\bar s} \, F \ .
\ee

With our choice of the streamlined nilpotent sector of the theory we need to identify the function $K_{s\bar s} (z^i, \bar z^{\bar i} )$ or its inverse $K^{s\bar s} (z^i, \bar z^{\bar i} )$ such that the standard supergravity expression for all chiral superfields, including the nilpotent one, in eq. \rf{VTMF},
is reduced to a given arbitrary bosonic potential depending on $n$ complex scalars $V(z^i, \bar z^{\bar i})$ when $s=0, \, F_s\neq 0$.  The simplest way to do it is to replace the general $F_s(z^i)= D_s W$ in eq. \rf{solution1} by a constant $F_s$ for our choice of $W= W_0 +   s \, F_s$ and to take into account that  $D_i W = K_i W_0$.  This gives  the following expression for the inverse \K metric:
\be
K^{s\bar s} (z^i, \bar z^{\bar i} ) =  {e^{-K(z^i, \bar z^{\bar i} ) } V(z^i, \bar z^{\bar i} ) + |W_0|^2 (3- K_i  K^{i\bar k} \bar K_{\bar k}) \over F_s \bar F_{\bar s}} \ .
\label{SolutionInv}\ee
The \K metric is identified as 
\be
K_{s\bar s} (z^i, \bar z^{\bar i} ) =  {F_s \bar F_{\bar s} \over e^{-K(z^i, \bar z^{\bar i} ) } V(z^i, \bar z^{\bar i} ) +  |W_0|^2 (3- K_i  K^{i\bar k} \bar K_{\bar k} ) } \ .
\label{Solution}\ee
Thus, the \K metric in streamlined supergravity depends only on $K(z^i, \bar z^{\bar i} )$ and its derivatives, and on $V(z^i, \bar z^{\bar i} )$.

{\it To summarize}: the streamlined supergravity depends on $n$ physical chiral superfields and one nilpotent superfield and has the following  \K potential and superpotential 
\be
K(z^i, \bar z^{\bar i},  s, \bar s)=K(z^i, \bar z^{\bar i} ) + {F_s \bar F_{\bar s}\over e^{-K(z^i, \bar z^{\bar i}) } V(z^i, \bar z^{\bar i})+ |W_0|^2 (3-K^{i \bar k} K_i K_{\bar k}) } \  
   s \,\bar s\, , \quad 
W= W_0 + F_s \, s \ .
\label{KW}\ee
We will sometimes refer to it as ``new'' rather than ``streamlined'' for simplicity.
The resulting bosonic action of  scalars is
\be
{ {\cal L} (z^i, \bar z^{\bar i})\over \sqrt{-g}} =  {R\over 2} + K_{i \bar k} (z^i,  \bar z^{\bar i})\, \partial  z^i \partial \bar z^{\bar k}+ K_{s\bar s}(z^i, \bar z^{\bar i})\, \partial  s \, \partial \bar {s}  - V(z^i, \bar z^{\bar i})  \ ,
\label{Baction}\ee
where the \K potential $K(z^i, \bar z^{\bar i})$ and potential $V(z^i, \bar z^{\bar i})$ are arbitrary. 

The full supergravity action, including the fermions, was presented in the unitary gauge for local supersymmetry $\psi^s=0$  in 
\cite{Kallosh:2015sea}, and a locally supersymmetric theory prior to gauge-fixing with a most general superpotential was given in \cite{Kallosh:2015tea}. In Appendix A, we will give a short summary of the complete streamlined supergravity with scalars, fermions, and vectors following  \cite{Kallosh:2015tea}. In Appendix B, we will describe the implementation of our methods to $SL(2,\mathbb{Z})$ invariant supergravity.

\section{Consistency issue of streamlined supergravity and its resolution}
The issue of consistency was raised in a particular example of streamlined/new supergravity in \cite{Yamada:2018nsk,Linde:2018hmx,Kallosh:2022vha} where it was observed that the expression for the \K metric $K_{s\bar s}$ of the kind we have in the general case in \rf{Solution} for certain values of model parameters is not positive definite at some points in the moduli space. This was revealed for some choice of cosmological parameters in the examples studied in these papers. If $S$  were a normal chiral multiplet, one would have to require that $K_{s\bar s}> 0$ to avoid a negative kinetic term for scalars and instability of the theory towards an infinite production of the corresponding particles.

An example given in \cite{Kallosh:2022vha} is for a model with one chiral multiplet and one nilpotent  multiplet:
\be
K(T, \bar T,  s, \bar s)=-3\a \ln(T+\bar T)+ \frac{F_s \bar F_{\bar s}}{(T+\bar T)^{3\alpha} V(T,\bar T) +3|W_0|^2(1-\alpha)}  \, s \,\bar s \ ,
\label{Kexample}\ee
In this case, the bosonic action following from this supersymmetric construction is 
\be
{ {\cal L} (T, \bar T)\over \sqrt{-g}} =  {R\over 2} - {3\alpha\over 4} \, {\partial T \partial \bar T\over ({\rm Re} \,  T )^2}-  V(T, \bar T)   \ .
\label{hyper2}\ee
The potential at some point in the moduli space $T_{min}$ may have a de Sitter minimum where it is equal to $\Lambda\ll 1$, and we find that at this point 
\begin{align}
K_{s\bar s}|_{T_{min}}=\frac{F_s \bar F_{\bar s}}{2^{3\alpha} \Lambda  +3|W_0|^2(1-\alpha)}\, .
\label{GXXm}\end{align}
The reason for the possible inconsistency is the fact that at the minimum
with a very small  cosmological constant $\Lambda$  only  models with $\a<1$ have a positive definite \K metric $K_{s\bar s}$ since
\be
K_{s\bar s}(T, \bar T)|_{T_{min}}  < 0 \qquad \rm{ if } \quad \a>1 \ .
\label{alpha}\ee
This was viewed for a while in \cite{Yamada:2018nsk,Linde:2018hmx,Kallosh:2022vha} as a restriction on the \K curvature of the moduli space, in this case, with $\mathcal{R}_K= -{2\over 3\a}$, the restriction was 
$
|\mathcal{R}_K| <{2\over 3}
$.
More recently, this issue was revisited in \cite{Kallosh:2025jsb}, taking into account that $S$ is a nilpotent chiral multiplet.  We will describe this case in a more general setting below. But first, we would like to discuss different points of view on the origin of non-linearly realized supersymmetry.

There was a trend, following \cite{Casalbuoni:1988xh,Komargodski:2009rz}, to obtain the nilpotent constraint in global supersymmetry models by starting with a linearly realized supersymmetry, giving a large mass to the scalar field, and integrating it out.  Analogous ideas were also developed in the context of supergravity. This procedure may be helpful for the reinterpretation of certain aspects of nonlinearly realized global supersymmetry. In particular, starting with linear supersymmetry models, one can evaluate the path integral in the limit of infinite sgoldstino mass \cite{Kallosh:2015pho} to derive de Sitter supergravity \cite{Bergshoeff:2015tra,Hasegawa:2015bza},
or investigate the decoupling limits to derive a model with a nilpotent multiplet; see, for example, \cite{Farakos:2013ih,DallAgata:2016syy}.  A consistency of this limiting procedure relies on the consistency of the theory with the linearly realized supersymmetry model before the formal limit to a nilpotent multiplet with nonlinear supersymmetry is taken.   This means, in particular, that one can establish a relation between supergravity with nonlinearly realized supersymmetry and with linearly realized supersymmetry only for the models with $K_{s\bar s}> 0$.

 However, we would like to remind here that supergravity theory with a nilpotent multiplet was derived from the underlying superconformal theory in \cite{Bergshoeff:2015tra,Hasegawa:2015bza,Kallosh:2015sea,Ferrara:2016een}.  This theory was not derived as a decoupling limit of supergravity with linearly realized supersymmetry, and its internal consistency does not rely on any relation between the linear and nonlinear realizations of supersymmetry in the large-mass limit. 

 Moreover, the existence of the D-branes shows that a non-linearly realized supersymmetry does not have to be a limit of the kind considered in  \cite{Casalbuoni:1988xh,Komargodski:2009rz}.   A supersymmetric Born-Infeld theory describing a D9-brane has a geometric form of the Volkov-Akulov-type \cite{Kallosh:1997aw}. There are Volkov-Akulov fermions living on the $\overline {D3}$ brane \cite{Bergshoeff:2015jxa}, which is a part of the KKLT mechanism of uplifting AdS vacua to dS vacua. The uplifting $\overline {D3}$ brane is either present or not, which means that non-linearly realized supersymmetry is either present or not; it was not constructed by a limiting procedure.
A recent review of these topics can be found in \cite{Dudas:2025ubq}.

 We will study below the issue of consistency of the supergravity with a nilpotent multiplet \cite{Bergshoeff:2015tra,Hasegawa:2015bza,Kallosh:2015sea,Ferrara:2016een} without attempting to re-interpret it in terms of the theory with the linear supersymmetry, where the decoupling limit is required.  We will find out that the sign of the  \K metric of the nilpotent multiplet does not matter: bosons and fermions of the nilpotent multiplet are not excited in the unitary gauge.  In this sense, the situation here is very similar to that in other gauge theories, where ghosts and tachyons that appear at intermediate stages of calculations are harmless if they disappear in the unitary gauge.

 To investigate this issue in detail,  we consider the full supergravity action, which includes the \K metric $K_{s\bar s}$, which could lead to an inconsistency because it is not positive definite. These are  kinetic terms in the action of the form
\be
K_{s\bar s} (\partial_\mu s \partial^\mu \bar s + \bar \psi^s \gamma_\mu D^\mu \psi^s) \ .
\label{kin}\ee
The nilpotent multiplet in supergravity satisfies the constraint  that the scalar  depends on the square of the spinor field 
\be
s= {(\psi^s)^2\over 2  F} \ .
\label{X} \ee
This makes the 1st term in the equation \rf{kin} quartic in spinor fields
\be
K_{s\bar s} \partial_\mu s \partial^\mu \bar s = K_{s\bar s} \partial_\mu \, {(\psi^s)^2\over 2  F} \partial^\mu \, 
{ (\bar \psi^{\bar s})^2\over 2F}\ .
\label{kin1}\ee
 According to \cite{Bellazzini:2016xrt}, this term is subject to positivity constraints, which would be inconsistent with the negative metric. Eqs. (5.3), (5.5)  in \cite{Bellazzini:2016xrt} show that in the case of global supersymmetry studied in  \cite{Casalbuoni:1988xh,Komargodski:2009rz,Bellazzini:2016xrt} indeed, one has to require that $K_{s\bar s}>0$.
The second term in eq. \rf{kin}  is quadratic in spinors, and its presence in the action directly indicates a possible inconsistency of the theory, as always in the case of the negative \K metric in quadratic kinetic terms.

 However,  in the case of a supergravity with a nilpotent multiplet (but not in  the case of a global supersymmetry),  we can use local supersymmetry to gauge-fix the fermion from the nilpotent multiplet to vanish, 
\be
\psi^s=0 \ .
\ee
 This is the unitary gauge, which was introduced in \cite{Bergshoeff:2015tra} in supergravity interacting with one nilpotent multiplet and in \cite{Kallosh:2015sea} in supergravity interacting with a nilpotent multiplet and other chiral multiplets. In this gauge, both terms in eq. \rf{kin}, with account of eq. \rf{X},  are absent from the gauge-fixed action, so the problem is solved. A negative \K metric for the nilpotent multiplet does not cause a physical instability of the system. This eliminates the corresponding constraints on the parameters of the cosmological theories of the kind we described in the example in eqs. \rf{Kexample}-\rf{alpha}. 
 
 Alternatively, one can take a different  unitary gauge $v=0$ where the goldstino $v$ is  interacting with the gravitino
 \be
 \bar \psi^\mu \gamma^\mu v= \bar \psi^\mu \gamma^\mu \Big [ {1\over \sqrt 2} e^{K(z^i, \bar z^{\bar i} )/2} ( \psi_z^i D_i W + \psi^s D_s W)\Big] \ .
 \ee
The requirement that goldstino $v$  vanishes in our new supergravity models means  that
\be
v= {1\over \sqrt 2} e^{K(z^i, \bar z^{\bar i} ) /2} (\psi^i_z K_i W_0 + \psi^s F_s)=0    \quad \Rightarrow \quad \psi^s= - {\psi^i_z K_i (z^i, \bar z^{\bar i} )W_0\over F_s} \ .
\ee
In the gauge where $v=0$, the fermion $\psi_s$ is replaced by a nonlinear function of moduli $(z^i, \bar z^{\bar i} )$ and a fermion  $\psi^i_z$ since  $K_i (z^i, \bar z^{\bar i} )$ does not have a moduli-independent term which would come from a \K potential linear in moduli $z^i$. But there is no reason to include the term $K(z, \bar z) \sim z+\bar z$, it will not even contribute to the metric $K_{z\bar z}$. 

Therefore, the kinetic terms of the boson $s$ and the fermion $\psi^s$ with a negative $K_{s\bar s}$ are not harmful. There is no vacuum instability due to $s$ and  $\psi^s$ since their kinetic terms are either absent in $\psi^s=0$ gauge, or in $v=0$ gauge involve terms with negative $K_{s\bar s}$ which are not quadratic in physical fermions $\psi^i_z$ since they contain coupling to moduli $z^i$.
Thus, {\it the streamlined supergravity defined by $ K_{s\bar s}$ in eqs. \rf{Solution} is consistent even in the models where $K_{s\bar s} <0$ at some points in moduli space.}

\section{Examples of streamlined supergravity models}
In all examples,  we use  a simple $z^i$-independent superpotential $W= W_0 + s\, F_s$ in eq. \rf{W} and present \K potentials and the bosonic action. The \K metric of the nilpotent field is always an example of the general expression in \rf{Solution}.
\subsection{ Models with hyperbolic geometry and an arbitrary potential}\label{Sec:hyper}

\subsubsection{E-models}


For models with an exponential approach to the plateau, consider the case of   E-models of $\alpha$-attractors  \cite{Ferrara:2013rsa,Kallosh:2013yoa}. In streamlined supergravity, these models are described by complex half-plane variables $T= e^{-\sqrt{2\over 3\a} \vp}+ i\, a $ in hyperbolic geometry with the nilpotent superfield $S$ with  $S^2=0$. 
The simplest models of this class have potentials with a plateau, which is approached exponentially fast at large $\vp$.
In particular, one may consider models  \rf{Kexample}, \rf{hyper2} with 
\be
V_{E}(T, \bar T)= \Lambda +V_0  (1- T)^{n} (1- \bar T)^{n}=  \Lambda + V_0\Big (1-e^{-\sqrt{2\over 3\a} \vp}\Big)^{2n}  \ .
\ee 
Here $T=\bar T=e^{-\sqrt{2\over 3\a} \vp} + i \, a$,  and inflation occurs at large values of the canonically normalized field $\vp$. To stabilize the axion field at $a = 0$, one may simply add to this potential a term $m^{2}(T,\bar T) (T -\bar T)^{2}$, which vanishes for $T= \bar T$.  (Note also that in this class of theories, one can construct consistent $\alpha$-attractor models with the same cosmological predictions without stabilizing the axion potential \cite{Achucarro:2017ing, Linde:2018hmx}).

These models have a potential with a plateau, which is approached exponentially fast at large $\vp$.
\be\label{eapp}
V_{E}(T, \bar T) = V_0 \Big(1- 2 n \, e^{-\sqrt{2\over 3\a} \vp} +\dots\Big) \ .
\ee 
The factor 2n in the last equation can be absorbed by a redefinition (a shift) of the field $\vp$.
This model can also be interpreted as pole inflation  \cite{Galante:2014ifa,Terada:2016nqg}, with the kinetic terms having a pole of second order at $ T = \bar T \to 0$ \rf{hyper2}.

\subsubsection{T-models}

One can consider inflationary models with hyperbolic geometry also in disk variables $Z$ related to half-plane variables $T$  by a Cayley transform, as shown in \cite{Cecotti:2014ipa,Kallosh:2015zsa}
\be
T= {1+Z\over 1-Z}\, ,  \qquad Z= {T-1 \over T+1} \ .
\label{cayley}\ee
One can show that 
 kinetic terms of chiral superfields $T$ and $Z$ are exactly the same, up to a Cayley transform in  \rf{cayley}
\be
{3\alpha\over 4} \, {\partial T \partial \bar T\over ({\rm Re} \,  T )^2}= {3\alpha} \, {\partial Z \partial \bar Z\over (1-Z\bar Z)^2} \ .
\ee
Note that these kinetic terms have a pole of the order $p = 2$.

The streamlined supergravity version of $\alpha$-attractors in disk variables, based on the general equation \rf{solution1}, corresponds to  
\begin{align}
K(Z, \bar Z, s, \bar s)=&-3\alpha\ln(1-Z\bar Z)+\frac{F_s \bar F_{\bar s}}{(1-Z\bar Z)^{3\alpha} \, V(Z, \bar Z) +3|W_0|^2(1-\alpha Z\bar Z)} s\bar s  \ .
\label{G}\end{align}
The bosonic action following from this supersymmetric construction is 
\be
{ {\cal L} (Z, \bar Z)\over \sqrt{-g}} =  {R\over 2} - {3\alpha} \, {\partial Z \partial \bar Z\over (1-Z\bar Z)^2}-  V(Z, \bar Z)   \ .
\label{hyper2Z}\ee 
Simplest examples include T-models of $\alpha$-attractors with $Z=\tanh {\varphi +i\vartheta \over\sqrt {6 \alpha}}$  \cite{Kallosh:2013yoa}. The potential with a heavy stabilized axion is
\be
V(Z,\bar Z)=\Lambda+  V_0 (Z\bar Z)^{n} + m^2(Z,\bar Z)\, (Z-\bar Z)^2 \ .
\ee
The inflaton potential for the canonically normalized inflaton field $\vp$ is the T-model potential
\be
V_{\rm T}=\Lambda +V_0 \tanh^{2n}{\varphi\over\sqrt {6 \alpha}} \ .
\ee
We assume that $\Lambda \ll V_{0}$, so one can ignore $\Lambda$ during inflation.
 The potential in disk variables with the stabilized axion and $Z=\bar Z =\tanh {\vp\over \sqrt{6\a}}$ during inflation at large $\vp$ is
\be
V(Z, \bar Z) =  V_{0 }\tanh^{2n}{\varphi\over\sqrt {6 \alpha}}=  V_0 \Big(1-2 n \, e^{-\sqrt{2\over 3\a} \vp} +\dots\Big) \ .
\ee
We find the same exponential approach to the plateau in terms of a canonical field $\vp$ as in the E-model \rf{eapp}. The factor 2n in the last equation can be absorbed by a redefinition (a shift) of the field $\vp$. This is a case of the pole inflation in disk variables, where the kinetic terms have a pole of the order $p = 2$ at the boundary $Z\bar Z \to 1$.

 The supergravity version of polynomial $\a$-attractors  \cite{Kallosh:2022feu} is also described in \rf{Kexample}, \rf{hyper2}, with the kinetic term with the pole of order $p = 2$. The potential in this case has particular features near the pole, where, e.g., for $k=2,4$ 
\be
V_{\rm pole}^{\rm ln}(T, \bar T) \to V_0 \Big(1-\Big({2\over \ln T \bar T}\Big)^{k}+\dots\Big)=  V_0 \Big(1- \Big ({\sqrt {2\over 3\a }\vp} \Big)^{-k}+\dots \Big) \ .
\ee
In this model, the plateau at large $\vp$ is approached polynomially \cite{Kallosh:2022feu,Kallosh:2025ijd}.

\subsection{\boldmath Polynomial approach to the plateau and pole inflation with $p > 2$ }

\subsubsection{\boldmath Pole inflation in half-plane variables}

We now describe pole inflation with kinetic terms that have a pole of order $p > 2$. To do it, we will take  the following function $K(T, \bar T, s, \bar s)$  defined in the general equation \rf{solution1}:
\be
K(T, \bar T, s, \bar s) = {  \mu^2\over 2} ( {T} \overline{T} )^{2-p\over 2}+ {F_s \bar F_{\bar s} \over {V}\left({T},\overline{{T}}\right) e^{-{\mu^2\over 2} (T \overline{T} )^{2-p\over 2} }+W_0^2 \left(3- {\mu^2\over 2} (T \overline{T} )^{2-p\over 2}\right)} \, s \bar{s}
\label{frac}\ee
When the axion is stabilized,  the bosonic action  at $T=\bar T$   following from this supersymmetric construction  is given by
\be
{ {\cal L} (T)\over \sqrt{-g}} =  {R\over 2} -{a_p\over 2}    \, {(\partial T)^2 \over T^{p}}- V(T)  \ , \qquad a_p= {\mu^2\over 4} (p-2)^2 \ .
\label{hyper2f}\ee
We may consider potentials near the pole of the kinetic term at $T\to 0$ of the form
\be
V|_{T=\bar T \to 0} \to V_0(1-T+\dots) \ .
\ee
This model describes pole inflation  \cite{Galante:2014ifa,Terada:2016nqg} with $p>2$, where in the canonical variables 
\be
V|_{\vp \to \infty} \to V_0(1-\Big({\mu\over \vp}\Big)^k+\dots)\ , \qquad k={2\over p-2} >0\ .
\ee

\subsubsection{\boldmath Pole inflation  in disk variables}

The streamlined supergravity version of pole inflation with pole order $p> 2$ in disk variables can be described by
\begin{align}
K(Z, \bar Z, s, \bar s)=& {  \mu^2\over 2} (1-Z \bar Z)^{2-p} +\frac{F_s \bar F_{\bar s}}{e^{{  \mu^2\over 2} (1-Z \bar Z)^{2-p} } \, V(Z, \bar Z) + |W_0|^2\Big(3-{  \mu^2\over 2} {(p-2)Z \bar Z (1-Z \bar Z)^{p-2}  \over 1+(p-2) Z \bar Z} \Big) }      s\bar s  \ .
\label{G2}
\end{align}
The bosonic action following from this supersymmetric construction is 
\be
{ {\cal L} (Z, \bar Z)\over \sqrt{-g}} =  {R\over 2} - {A_{p}\over 2} \, {\partial Z \partial \bar Z\over (1-Z\bar Z)^p}-  V(Z, \bar Z)   \ , \qquad A_{p}={  \mu^2} (p-2)\big(1+(p-2)Z\bar Z\big) \ .
\label{hyper2Z2}\ee 
This is the pole-inflation model in which the kinetic terms have a pole of order $p > 2$ at the boundary $Z\bar Z \to 1$. The polynomial approach to the plateau in terms of the canonical field $\vp$ is similar to the one in the model \rf{frac}. Observational results of these inflationary models depend only on $p$, on the value of the potential  $V(Z, \bar Z)$ at $Z\bar Z \to 1$, and on the value of $A_{p}$ at $Z\bar Z \to 1$.
\be
a_{p} = A_{p} \Big |_{Z\bar Z \to 1} =  \mu^2  (p-2)(p-1) \ .
\ee

\section{Discussion} 

In this paper we presented an efficient and relatively simple version of the streamlined $N=1$ supergravity with chiral multiplets and a nilpotent multiplet $S^2=0$ with the general type  \K potential  $K(z^i, \bar z^{\bar i} )$ leading to  {\it any desired scalar potential } $V(z^i, \bar z^{\bar i})$,  once the proper choice of the \K metric of the nilpotent multiplet is made. The total \K potential,  superpotential, and the bosonic action of this new supergravity are given in eqs. \rf{KW}, \rf{Baction}.
We provided some examples illustrating our methods in applications to inflationary potentials.

In Appendix A, we present a complete streamlined supergravity including scalars, vectors, fermions, and one nilpotent chiral multiplet with a  general superpotential 
 $
W (z^i, s)= g(z^i) + s \, f(z_i)
$ as constructed in  \cite{Kallosh:2015tea}. 
 The vector part of the theory 
and a holomorphic gauge kinetic function 
$
f_{AB} (z^i, s) = f_{AB0}  (z^i) + s \,   f_{ABs}(z^i) 
$ and momentum maps ${\cal P}_A$
 correspond to a standard supergravity, as given in the book  \cite{Freedman:2012zz}. We also have a comment there about the work in  \cite{Cribiori:2017laj} concerning Fayet-Iliopoulos terms in supergravity
without gauged R-symmetry.

The major difference with the book \cite{Freedman:2012zz} before gauge-fixing local supersymmetry is in the fermionic sector of the theory, as one can see in Eq. \rf{eq:L10}. In the unitary gauge where the fermion of the nilpotent multiplet $\psi^s$ vanishes, all extra terms vanish. The resulting theory is given in Eq. \rf{unit} for physical and nilpotent scalars, vectors, and fermions, where one has to put the nilpotent scalar to zero. This last step leaves the additional part in the scalar potential, $e^K F_s K^{s \bar s} \bar F_{\bar s}$, relative to the standard supergravity without the nilpotent multiplet. This function is an arbitrary function of $(z^i, \bar z^{\bar i})$. 

In Appendix B, we apply this general method to $SL(2,\mathbb{Z})$ invariant supergravity theories.

We conclude that in the framework of streamlined supergravity, it is easy to find de Sitter solutions, to stabilize moduli, and to embed in supergravity many phenomenologically interesting bosonic models with all possible sets of $K(z^i, \bar z^{\bar i} )$ and $V(z^i, \bar z^{\bar i})$.

\section*{Acknowledgement}
We acknowledge many stimulating discussions over the years on supergravity with nilpotent multiplets with I. Antoniadis, J.J. Carrasco, S. Ferrara, D. Freedman, E. McDonough, D. Roest, M. Scalisi, A. Van Proeyen, T. Wrase, and Y. Yamada.   This work was supported by the Leinweber Institute for Theoretical Physics at Stanford and by NSF Grant PHY-2310429.

\appendix

\section{Complete streamlined supergravity with scalars, fermions, and vectors }

The explicit supergravity action describing the interaction of supergravity with an arbitrary number of chiral and vector multiplets, as well as one nilpotent chiral multiplet, was presented in \cite{Kallosh:2015tea}. It is defined by the \K potential of the form given in eq. \rf{K}, by the superpotential 
\be
W (z^i, s)= g(z^i) + s \, f(z_i) \ ,
\label{Wz}\ee
by a holomorphic gauge kinetic function $f_{AB}$ 
\be
f_{AB} (z^i, s) = f_{AB0}  (z^i) + s \,   f_{ABs}(z^i) \ ,
\label{gkin}\ee
and real scalar momentum maps $\mathcal{P}_A$  defining the gauge transformations of the chiral multiplet scalars.
We use a simpler $W$ when defining the streamlined supergravity above, with $g(z^i)=W_0$ and $f(z_i)=F_s$. But we have not yet included fermions and vector multiplets with the coupling defined in $\rf{gkin}$; we will do so now.

The complete supergravity action for the choice of the \K potential in \rf{K}, choice of $W$ in \rf{Wz}, choice of $f_{AB}$ in  \rf{gkin} was given in \cite{Kallosh:2015tea} in eqs. (2.7) - (2.8)  where $e^{-1} {\cal L} ^{\rm book}$ are presented in detail in equations (18.6) - (18.19) in the book  \cite{Freedman:2012zz}. We reproduce the result in our notation here (where $s=z^1$)
\bea\label{eq:L10}
e^{-1} { \cal L}_{\rm final} = \Big [ e^{-1} { \cal L}^{\rm book} \Big ]_{s=\frac{(\psi^s)^2}{2 F^s}} -   \frac{(\psi^s)^2\, \, }{ 4 K_{s \bar s} (F^s\bar F^{\bar s})^2} \left|  K_{s\bar s} \frac{\Box(\psi^s)^2}{2 F^s}+ B^{s}  \right |^2 \,\,\,
\eea
with
\bea\label{eq:B1}
F^s \equiv  K^{s \bar s} \lp -\rme^{\frac{K}{2}}\,\overline{W}_{\bar s} +{1\over 2} (\partial_i K_{s\bar s}) \bar \psi^s \psi^i+\frac{1}{4 } \bar f_{AB \bar{s}} \bar{\lambda}^A P_R \lambda^B \rp\,,  \quad B^{s}  \equiv  e^{-1} { \delta { \cal L}^{\rm book}\over \delta \bar s} |_{s= \bar s=0}\, .
\eea
Note that the last term in equation \rf{eq:L10} already has the maximal power of the (undifferentiated) spinor $\psi^s$, so that we can drop all terms in $B^s$ that contain $\psi^s$ or $\bar \psi^{\bar s}$. Explicit expressions for  $B^s$ for some examples are given in \cite{Kallosh:2015tea}.

This complete action, including chiral and vector multiplets, is complicated by local supersymmetry before gauge fixing. It is complicated even in the case of pure de Sitter supergravity \cite{Bergshoeff:2015tra}. But in the unitary gauge with $\psi^s=0$, both the action of the pure de Sitter supergravity in \cite{Bergshoeff:2015tra} as well as the action with chiral and vector multiplets become very simple. In our case here, we find in the unitary gauge with $\psi^s=0$
\bea\label{unit}
e^{-1} { \cal L}_{\rm final} = \Big [ e^{-1} { \cal L}^{\rm book} \Big ]_{s=0, \, \psi^s=0} 
\label{final}\eea
In particular, we have a standard potential for all chiral multiplets as well as a leftover $V^{\rm book}_{s,\bar s} 
$ from the presence of the nilpotent multiplet, since $D_s W =F_s \neq 0$
\be
V|_{s=0}= V^{\rm book}_{\rm standard} + V^{\rm book}_{s,\bar s} = e^K ( D_i W K^{i\bar k} \bar D_{\bar k} \bar W - 3 W\bar W) +e^K (F_s K^{s\bar s} \bar F_{\bar s}  ) \ .
\label{Vfinal}\ee
Finally, we would like to note here that  particular choice of  $K_{s\bar s}$ and $F_s$ in  our scalar potential in \rf{VTMF}  
 $e^K |F_s|^2 K^{s \bar s}$ reproduces the potential  (1.1) in  \cite{Cribiori:2017laj}. Namely, with $K_{s\bar s}= e^{K/3}, |F_s|^2= {\xi^2\over 2}$,  we find that our potential is equal to ${\xi^2\over 2} e^{2K\over 3}$. This potential was interpreted in \cite{Cribiori:2017laj} in the context of the gauge-fixed superconformal theory as a Fayet-Iliopoulos term in supergravity without gauged R-symmetry. It was, however, also observed there that ,in the context of constrained multiplets, this potential originates from the superpotential $W= W(z)- {\xi\over \sqrt 2} s$. This is in agreement with our term F-term potential $e^K |F_s|^2 K^{s \bar s}$ with  $K_{s\bar s}= e^{K/3}$, $|F_s|^2= {\xi^2\over  2}$.
 
 \section{\boldmath $SL(2,\mathbb{Z})$ invariant supergravity}
 
 We consider supergravity action with a physical chiral superfield where its scalar $\tau= iT$ is a modulus of the upper half-plane, and there is a nilpotent superfield with a scalar $s$ and fermion $\psi^s$. We study the action, including fermions, gravitino, and the fermion partner of  $T$, which we call  $\psi^T$,  in the unitary gauge  $\psi^s=0 $. According to eq. \rf{final}, this means that we have to start with standard supergravity with 2 chiral superfields with scalars $T, s$, and take into account that the potential \rf{Vfinal} differs from the standard supergravity potential with only one superfield $T$.
 
Using the streamlined supergravity models with $W=W_0+ s\, F_s $, one can construct a consistent supergravity embedding of the models with any potential $V(z^i, \bar z^{\bar i})$. This method made it possible to construct a large set of supergravity models with $SL(2,\mathbb{Z})$- invariant potentials; see Sect. 9 of  \cite{Kallosh:2024ymt}. The \K invariant   functional $G$ there is
 \be
 G(T, \bar T; s, \bar s)=  -3\a \log (T +\bar T )+ \log |W_0+ s\, F_s |^2 +G_{s\bar s}(T, \bar T)  s\bar s \ ,
 \ee
 where
 \be
G_{s\bar s}= K_{s\bar s} = {|F_s|^2\over (T+\bar T)^{3\a} V(T, \bar T) +3|W_0|^2 (1-\a)}  \ .
 \ee
The corresponding bosonic action is
\be
{{\cal L}(T, \bar T)\over \sqrt{-g}}= {R\over 2} -{3\a\over 4} {\partial T \partial\bar T\over ({\rm Re} T)^2}- V(T, \bar T) \ .
\ee 
Here, the scalar kinetic term is $SL(2,\mathbb{Z})$ invariant. Therefore, if the potential  $V(T, \bar T)$ is   $SL(2,\mathbb{Z})$ invariant, the total bosonic part of the supergravity action is  $SL(2,\mathbb{Z})$ invariant.   It was sufficient here to use the simplest superpotential of the streamlined supergravity  $W=W_0+ s\, F_s $.

 However, if we also want to have the fermion part of the action with gravitino and $\psi^T$
   to be $SL(2,\mathbb{Z})$ invariant, we should use a more general $T$-dependent superpotential of the kind given in eq. \rf{Wz}. In supergravity with one chiral multiplet without the nilpotent one, the $SL(2,\mathbb{Z})$ invariant theory, including gravitino and $\psi^T$ was presented in \cite{Ferrara:1989bc}. The proof of $SL(2,\mathbb{Z})$ invariance there required that the generating functional 
 \be
 G(T, \bar T) = -3\a \log (T +\bar T )+ \log W(T)  + \log \bar W(\bar T ) 
\label{GT}\ee
 should be invariant under the $SL(2,\mathbb{Z})$ transformations. The   superpotential $W(T)$ must transform as a modular form of weight $-3\a$ to compensate the $SL(2,\mathbb{Z})$ transformation of $K(T, \bar T) =-3\a \log (T +\bar T )$. The gravitino, which is neutral with respect to the $SL(2,\mathbb{Z})$ transformations, couples to  $G(T, \bar T)$.  
\be
e^{G(T, \bar T)\over 2} \bar \psi_{\mu R}  \sigma^{\mu\nu} \psi_{\nu R} +h.c. 
\ee 
The fermion superpartner of $T$, the field 
 $\psi^T$,  transforms as a modular form of weight $-2$ and it  couples to $ G(T, \bar T)$ and its derivatives. This is why the  $SL(2,\mathbb{Z})$ invariance of $G(T, \bar T)$  is necessary and sufficient for the invariance of the fermion part of the action in the case studied in \cite{Ferrara:1989bc}.
 
Now, as different from  \cite{Ferrara:1989bc},  we begin with an action containing two multiplets: the physical multiplet and the nilpotent multiplet. Gravitino and $\psi^T$ are coupled to 
 $G(T, \bar T; s, \bar s)$ and its $T$ derivatives, where
$
G(T, \bar T; s, \bar s)= G(T, \bar T) + \dots
$
Here $\dots $ means terms linear in $s$ and $\bar s$ and in $s\bar s$. But according to eq. \rf{final} in the unitary gauge $\psi^s=0$, we have to take the standard action for both multiplets at $\psi^s=0, s=0$. Thus, in the final action, gravitino and $\psi^T$ are coupled to  
$
G(T, \bar T; s, \bar s)\Big |_{s=0} = G(T, \bar T) 
$ 
and its derivatives in $T, \bar T$ as shown in eq. (7) in \cite{Ferrara:1989bc}.

To construct this $SL(2,\mathbb{Z})$ invariant supergravity with one physical multiplet and one nilpotent multiplet, we proceed as follows.
We choose a  superpotential 
\be
W (T, s)= W(T) + s \, F_s \ ,
\label{Wsl}\ee
where $W(T)$ is a modular form of weight $-3\a$. After that, our generating potential \rf{GT} becomes $SL(2,\mathbb{Z})$ invariant.
In this superpotential, the constant $W_0$ is replaced by a function $W(T)$, but we still keep the constant $F_s$. 
Our generating functional is
  \be
 G(T, \bar T; s, \bar s) = -3\a \log (T +\bar T )+ \log (W(T) +s F_s) + \log( \bar W(\bar T ) +\bar s \bar F_s)+K_{s\bar s} (T, \bar T)s\bar s  
\label{GTS}\ee
 As we explained above, in the unitary gauge the physical fermions couple only to $G(T, \bar T; s, \bar s))\Big |_{s=0}$, which is an expression given in eq. \rf{GT}. Therefore, the total supergravity action, bosonic and fermionic part,  is $SL(2,\mathbb{Z})$ invariant under the condition that the potential $V(T, \bar T)$ is $SL(2,\mathbb{Z})$ invariant and eq. \rf{Vfinal} is satisfied. We therefore find that for any choice of the  $SL(2,\mathbb{Z})$ invariant potential $V(T, \bar T)$ the inverse metric of the nilpotent superfield is
 \be
 K^{s\bar s} (T, \bar T ) ={(T+\bar T)^{3\a} V(T, \bar T) - D_T W(T) K^{T \bar T} \bar D_{\bar T} \bar W(\bar T) + 3 W(T)\bar W(\bar T) \over F_s  \bar F_{\bar s} }  \ .
\label{VTMF1}\ee
 The total \K potential is 
\be
K(T, \bar T,  s, \bar s)=-3\a \log (T+\bar T) + {F_s \bar F_{\bar s}\over(T+\bar T)^{3\a} V(T, \bar T) - D_T W(T) K^{T, \bar T} \bar D_{\bar T} \bar W(\bar T) + 3 W(T)\bar W(\bar T) }  
   s \,\bar s \nonumber 
\label{KWtau}\ee
and the total superpotential is given in Eq. \rf{Wsl}. These expressions for $K(T, \bar T,  s, \bar s)$ and $W (T, s)$ define the supergravity theory, which is $SL(2,\mathbb{Z})$ invariant.

\bibliographystyle{JHEP}
\bibliography{lindekalloshrefs}
\end{document}